\documentclass[12pt]{article}
\usepackage{fullpage}
\usepackage{authblk}
\usepackage{amsmath, amsthm, amsfonts,  amssymb, latexsym}
\usepackage{amsmath,amsfonts,latexsym,amssymb,amscd}
\usepackage[latin1]{inputenc}
\usepackage[T1]{fontenc}
\usepackage{pspicture}
\usepackage{verbatim,amsthm,curves,graphics}
\usepackage{mathrsfs}

\def\ben{\begin{equation}}

\def\een{\end{equation}}

\def\bena{\begin{eqnarray}}

\def\eena{\end{eqnarray}}

\def\f(#1/#2){\frac{#1}{#2}}

\def\Frac(#1/#2){\left(\frac{#1}{#2}\right)}

\def\chris(#1-#2-#3){{\mit \Gamma}^{#1}{}_{{#2}{#3}} }

\def\tilchris(#1-#2-#3){\tilde{{\mit \Gamma}}^{#1}{}_{{#2}{#3}}}

\def\hatchris(#1-#2-#3){\hat{{\mit \Gamma}}^{#1}{}_{{#2}{#3}}}

\theoremstyle{definition}

\renewcommand{\epsilon}{\varepsilon}

\newcommand{\mc}{{\mathbb C}}

\newcommand{\M}{{\bf M}}

\renewcommand{\O}{{\mathcal O}}

\newcommand{\A}{{\mathcal A}}

\begin{document}

\title{The Operator Product Expansion in Quantum Field Theory}

\author{Stefan Hollands}
\affil{\small Institute for Theoretical Physics, Leipzig University, Br\" uderstrasse 16, 04103 Leipzig, and MPI-MiS, Inselstrasse 22, 04103, Leipzig, Germany, stefan.hollands@uni-leipzig.de}
\author{Robert M. Wald}
\affil{\small Enrico Fermi Institute and Department of Physics,
The University of Chicago, 933 East 56th Street, Chicago, Illinois 60637, USA, rmwa@uchicago.edu}

\date{\today}

\maketitle

\begin{abstract}
Operator product expansions (OPEs) in quantum field theory (QFT) provide an asymptotic relation between products of local fields defined at points $x_1, \dots, x_n$ and local fields at point $y$ in the limit $x_1, \dots, x_n \to y$. They thereby capture in a precise way the singular behavior of products of quantum fields at a point as well as their ``finite trends.'' In this article, we shall review the fundamental properties of OPEs and their role in the formulation of interacting QFT in curved spacetime, the ``flow relations'' in coupling parameters satisfied by the OPE coefficients, the role of OPEs in conformal field theories, and the manner in which general theorems---specifically, the PCT theorem---can be formulated using OPEs in a curved spacetime setting.
\end{abstract}

\section{Products in classical versus quantum field theory}
\label{sec:intro}

Classical field theories are an extremely successful class of physical models describing a wide range of phenomena ranging from fluids to electromagnetism and general relativity. Classical fields are normally described by smooth functions, so their values and the values of their derivatives are well defined at each spacetime point $x$. Classical field theories are normally formulated in terms of partial differential equations (PDEs) that provide relationships between a finite number of partial derivatives of the physical fields 
at each point $x$ in spacetime.
For smooth fields, no difficulties arise in taking products of fields and their derivatives at $x$, as needed to write down any nonlinear PDE. Even in situations where the solutions develop irregularities such as shocks, or when developing an abstract mathematical solution theory in terms of function spaces with weak regularity, there is normally no problem in defining the relationships between the fields and their derivatives 
at a point needed to write down the PDE in question.

The situation in quantum field theory (QFT)---and to a varying degree also in classical field theories with random drivers (see e.g. \cite{Hairer})---is markedly different in this regard because the fluctuations of the field evaluated at a sharp spacetime point $x$ are typically divergent. The fields only become well defined as distributions, i.e., when one suitably averages them over a spacetime region. Consequently,
one cannot in any straightforward manner define the product of quantum fields---such the square $\phi(x)^2$---at point $x$. It is therefore very far from  evident 
how to write down meaningful PDEs for quantum fields analogous to those in classical field theories, and therefore how to ``define'' the theories in the first place!

In quantum field theory, one obtains infinite values when one attempts to straightforwardly calculate quantities like the expectation value $\langle \Psi | \phi(x)^2 | \Psi \rangle$ of the square of a quantum field $\phi(x)$ at point $x$ in state $|\Psi \rangle$. This
simply reflects the fact that the fluctuations of $\phi(x)$ in the given state are infinite, i.e., the variance of individual measurements of $\phi(x)$ in that state would be infinite. On the other hand, 
the operator that one would {\it really} like to define is one whose classical counterpart is $\phi(x)^2$. In order to obtain such a field $\phi^2$, one must find a way of discarding the fluctuations and retaining a suitable finite part, thought of as defining the general larger scale {\it trend} expressed by the quantity corresponding to the classical field $\phi(x)^2$, as opposed to the extremely short scale {\em fluctuations} of the quantum field $\phi(x)$. It is this finite part that should appear in the PDEs or other relations defining the theory. Similar issues arise for the fluctuations of stochastic origin occurring, e.g., in Brownian motion.

\section{Operator product expansion for a free Klein-Gordon Field}
\label{sec:outline}

The operator product expansion (OPE) provides a mathematically precise characterization of the fluctuations of quantum fields on small scales. The formulation of the OPE was due originally to \cite{wilson,zimmermann}. In this section, we will consider the OPE of a free massless Klein-Gordon (KG) quantum field. This will serve to illustrate the nature of OPEs and tie in OPEs with the discussion of the previous section.

Consider a free massless KG quantum field in 
4-dimensional Minkowski spacetime with 
``basic'' KG field $\phi(x)$. The OPE for the product of the field at two different points $x_1$ and $x_2$ has the form (with distributional $\epsilon \to 0+$ prescription understood)
\ben
\label{KGeq}
\phi(x_1) \phi(x_2) \approx \frac{1}{2\pi^2} \frac{1}{(x_1-x_2)^2 +i\epsilon(x_1^0-x_2^0)} {\bf 1} + \mathcal{O}(y) + \dots 
\een
where $\bf{1}$ denotes the identity operator, $\mathcal{O}$ represents a new local field of the theory, and the terms indicated by ellipses contain other new local fields at $y$ but whose coefficients vanish in the limit as $x_1$ and $x_2$ approach point $y$. The ``$\approx$'' symbol (as opposed to an ``$=$'' sign) in eq.~(\ref{KGeq}) is intended to indicate that (\ref{KGeq}) describes an asymptotic expansion of $\phi(x_1) \phi(x_2)$, valid in the limit as $x_1, x_2 \to y$. This asymptotic expansion is meant to hold for the expectation value of both sides of (\ref{KGeq}) in a state $|\Psi \rangle$. However, although it can be verified that this asymptotic expansion is satisfied by expectation values in the vacuum state $|0 \rangle$ or in $n$ particle states with smooth mode functions, one can find states in the standard Fock space of the free Klein-Gordon field---such as $n$ particle states with singular mode functions---whose expectation values do not satisfy (\ref{KGeq}). Thus, (\ref{KGeq}) may be viewed as imposing a ``regularity condition'' on states analogous to smoothness conditions imposed on classical fields.

Note that the coefficient of $\bf{1}$ diverges as  $x_1-x_2 \to 0$. This corresponds to the infinite fluctuations of $\phi(x)$ alluded to in the previous section. In this example, these infinite fluctuations are proportional to the identity operator, expressing the fact that the divergence of the fluctuations is the same in any (regular)
state $|\Psi\rangle$. The field $\mathcal{O}$ expresses the ``trend'' of $\phi(x)^2$ once one has removed these infinite fluctuations. We take this field $\mathcal{O}$ to be the {\it definition} of the square 
of $\phi(x)$, i.e., we define $\phi^2=\mathcal{O}$. Compared to the naive definition of $\phi(x)^2$, this definition has removed the ``infinite fluctuations'' that would otherwise swamp the general ``trend'' of $\phi(x)^2$. The terms indicated by ellipses are irrelevant for the purpose of defining the quantum field $\phi^2$, but they would be relevant, e.g., for defining the square of derivatives of $\phi$. 

The prescription we have just given for defining $\phi^2 = \O$ corresponds to the so-called point-spliting renormalization prescription, see e.g. \cite{hw1,hw2}. For other observables, such as $\partial_a \phi \partial_b \phi$, that are quadratic in the quantum field $\phi$, one would similarly consider the OPE of $\partial_a\phi(x_1) \partial_b\phi(x_2)$, which can be obtained by differentiating the OPE \eqref{KGeq}. One would then extract $\partial_a \phi \partial_b \phi$ as the first finite term appearing on the right side of this OPE. In particular, the (renormalized) stress-energy tensor $T_{ab}$ of the Klein-Gordon field can be obtained in this manner.

The generalization of (\ref{KGeq}) for a free KG quantum field obeying $\nabla^a \nabla_a \phi=0$ in a Lorentzian curved spacetime is given as follows. In curved spacetime, the coefficient of the identity---which describes the most divergent fluctuations---is obtained by replacing the flat spacetime expression by the covariant expression
\ben
\frac{1}{2\pi^2} \frac{1}{(x_1-x_2)^2+i\epsilon(x_1^0-x_2^0)} \to 
\frac{1}{2\pi^2}\left( \frac{\Delta_{12}^{1/2}}{s_{12}+i\epsilon \tau_{12}} + \dots \right) 
\label{KGcur}
\een
where $s_{12}$ is the (signed) squared geodesic distance between $x_1$ and $x_2$, $\Delta^{1/2}_{12}$ is the Van-Vleck determinant (which measures the change in the area-cross section of a bundle of geodesics) and $\tau_{12}$ stands for a time difference between $x_1$ and $x_2$ measured with an arbitrary global time function $\tau$. The dots $...$ represent less divergent terms (in particular, terms with $\ln s_{12}$ dependence) that are constructed from the spacetime metric in a local and covariant manner.

As in flat spacetime, the quantity $\phi^2 = \O$ would be extracted as the first finite term in the OPE of $\phi(x_1) \phi(x_2)$. However, in curved spacetime, ambiguities arise in the choice of the identity coefficient (\ref{KGcur}), since ``divergent quantities'' are inherently ambiguous up to addition of ``smooth quantities.'' Such ambiguities are also present in flat spacetime but can be resolved by requiring that the coefficient of $\bf{1}$ in eq.~(\ref{KGeq}) be equal to $\langle 0 | \phi(x_1) \phi(x_2)| 0 \rangle$, where $|0 \rangle$ denotes the Poincare invariant vacuum state. However, there are no such ``preferred states'' on a general curved spacetime, so the choice of the coefficient of the identity right side of eq.~(\ref{KGcur}) is inherently ambiguous up to addition of a smooth, symmetric function $f(x_1, x_2)$ that is locally and covariantly constructed from the spacetime metric and has appropriate scaling behavior. This leads to an ambiguity in the definition of $\phi^2$ corresponding to $\phi^2 \to \phi^2 + c R \bf{1}$, where $c$ is an arbitrary constant and $R$ denotes the scalar curvature. Similar ``local curvature ambiguities'' arise for the stress-energy tensor, although in this case the form of the ambiguous terms is further restricted by the requirement that $T_{ab}$ satisfy $\nabla^a T_{ab} = 0$.

We have discussed above the OPE of the product $\phi(x_1) \phi(x_2)$ of two ``basic fields''. More generally, there is a similar OPE of a product of $n$ basic fields $\phi(x_1) \cdots  \phi(x_n)$. This OPE will contain divergent terms proportional to the identity operator $\bf{1}$ and/or $\phi^k$ with $k < n$. The first finite term in this OPE can be used to define $\phi^n$. Similarly, the OPE of derivatives of $\phi(x_1) \cdots  \phi(x_n)$ can be used to define $n$-fold products of derivatives of $\phi$. 

Much more generally, we can consider the collection of all local operators $\mathcal{O}_i$ in the theory, i.e., the ``basic field'' $\phi$ together with the ``composite fields'' $\phi^n$ and $n$-fold products of derivatives of $\phi$. The collection of all the local fields satisfies OPEs of the form
\ben
\label{OPE}
\mathcal{O}_{1} (x_1) \cdots \mathcal{O}_{n}(x_n) \approx \sum_k C_{1 \dots n}^k(x_1, \dots, x_n; y) \mathcal{O}_k(y), 
\een
where the ``OPE coefficients'' $C_{1 \dots n}^k(x_1, \dots, x_n; y)$ are distributions whose explicit form can be found in \cite{KW}.

In summary, we have illustrated above for the case of a free, massless KG field that the OPE plays two distinct key roles in quantum field theory: (1) It provides a means for discarding the infinite fluctuations of a quantum field so as to obtain quantum observables that correspond to classical nonlinear field observables. This enables one to then impose PDEs or other nonlinear relations on the quantum field theory. (2) It provides a regularity condition on states.

\section{Axioms for OPEs of nonlinear QFTs}
\label{sec:axioms}

Quantum field theories in which the basic quantum field satisfies linear equations of motion---such as the KG field considered in the previous section---are mathematically well defined. However, apart from perturbative expansions about a linear theory, certain very special theories in low dimension, and certain conformal field theories (see section \ref{sec:cft}), the definition of nonlinear quantum field theories remains a fundamental open issue. Most approaches toward the formulation
of QFT in Minkowski spacetime make heavy use of the existence of a Poincare invariant vacuum state. In axiomatic settings such as the Wightman axioms \cite{streaterwightman}, the content of the theory is expressed in terms of the $n$-point functions of the quantum field(s) in the vacuum state, and the notion of Poincare invariance plays a crucial role when formulating fundamental properties of the $n$-point functions. 
In the path integral formulation \cite{glimm}, it enters the precise definition of the path space (``contours'') that are to be considered.

However, our universe is not isometric to Minkowski spacetime, and the formulation of a quantum field theory would be highly deficient if it could only be formulated in some very special spacetimes. We cannot expect a QFT to be formulated in spacetimes with such bad causal behavior that the corresponding classical theory is ill posed. However, it would be appropriate to demand that a QFT can be formulated in all globally hyperbolic spacetimes, since classical field theory is well posed in such spacetimes. However, 
in a general globally hyperbolic curved spacetime, there are no spacetime symmetries, and correspondingly, no preferred states on which one could naturally base axiomatic formulations or path integral approaches. For this reason, we do not believe that the formulation of quantum field theory should be based on the existence of some ``preferred vacuum state.''

For a linear quantum field in a general curved spacetime with a non-compact Cauchy surface (i.e., an ``open universe''), not only will there be no ``preferred vacuum state'' but there also will be no preferred Hilbert space construction of the theory, i.e., there will be infinitely many unitarily inequivalent constructions and no obvious way, in general, of selecting a ``distinguished'' construction. The situation surely will not be better for nonlinear quantum fields. This necessitates adopting the algebraic viewpoint on QFT \cite{haag}. In the algebraic viewpoint, the fields on a spacetime are viewed as elements of an abstract $*$-algebra, $\A$. States are defined as linear, maps $\omega: \A \to \mc$ 
that are positive in the sense that
$\omega(A^* A) \ge 0$ for
all $A \in \A$, where $\omega(A)$ is interpreted as the expectation value of the observable $A$ in the state $\omega$.
Given a state $\omega$, the GNS construction (see e.g., \cite{haag}) produces a Hilbert space representation of the algebra, so one can recover all results one would obtain in a Hilbert space approach. The main advantage of the algebraic approach is that it allows one to simultaneously consider all states arising in all Hilbert space constructions, without having to select a preferred construction at the outset.

In the algebraic approach, one usually assumes some $C^*$ (or stronger) structure of the algebra of observables. A model for this is the free KG field, where quantities formally corresponding to the ``bounded quantity'' $e^{i \phi(f)}$ (where $\phi(f)$ denotes the quantum field $\phi$ smeared with a test function $f$) generate an algebra---known as the Weyl algebra---with a natural $C^*$ structure. This construction works because the commutator of free fields is proportional to $\bf{1}$, so the field commutation relations can be expressed as product relations between the Weyl generators. However, in a nonlinear field theory it is far from obvious how to define corresponding bounded quantities in such a way that the relations one wishes to impose on the quantum field can be expressed as algebraic relations on the bounded quantities. Indeed, the relations of the quantum field theory would be most naturally expressed as local relations on the ``unbounded'' local quantum fields themselves. However, such unbounded fields could not be expected to have more than a $*$-algebra structure. 

Our view is that the formulation of a quantum field theory should be based on it's OPE, i.e., that the OPE should be elevated to a fundamental structure of QFT rather than merely a convenient tool to, e.g., define ``composite fields,'' as described in the previous section. There is considerable evidence that all QFTs of interest will have OPEs of the form (\ref{OPE}). In particular, such OPEs exist order-by-order in perturbation theory for renormalizable theories \cite{hollandsQFTCST, kopper, zimmermann}, and, as will be discussed in \ref{sec:cft}, they exist in conformal field theories. Furthermore, as argued in \cite{HollandsWaldOPE}, essentially\footnote{We say ``essentially'' because it does not appear possible to impose certain relations---such as commutation/anti-commutation relations at large spacelike separations---via OPEs. Such additional relations would have to be appended to the formulation of the theory.} all of the information about a quantum field theory---such as its equations of motion---can be formulated in terms of its OPEs. Thus, we believe that a quantum field theory should be viewed as being defined by specifying its OPEs, i.e., by providing a complete list of its OPE coefficients $C_{1 \dots n}^k(x_1, \dots, x_n; y)$. Of course, these OPE coefficients must satisfy a number of properties in order to define an acceptable quantum field theory. We refer to these properties as ``axioms for OPEs'' and, as reviewed below, we proposed a list of such axioms in \cite{HollandsWaldOPE}. Our view is that a quantum field theory corresponds to a specification of OPE coefficients $C_{1 \dots n}^k(x_1, \dots, x_n; y)$ that satisfy these axioms.

Before describing these axioms, we make a few preliminary remarks. An important aspect of our formulation is that the OPE coefficients
are required to be given for any globally hyperbolic $(M,g_{ab})$, as well as any choice of orientation, $o$, and time orientation, $\tau$, on spacetime, as well as, if relevant, a spin-structure and/or other choice of principal fibre bundle structure, e.g., in the case of gauge theories. In particular, we do not allow ``exceptional'' QFTs that for some reason can only be defined, e.g., on Minkowski spacetime due to its special structure. 

The local field observables $\O_i$ that appear in the OPE (\ref{OPE}) are abstract quantities that will become operator-valued distributions in Hilbert space representations of the theory.
We assume that there is a ``hermitian conjugation operation'' $\dagger$ acting on the fields $\O_i$ that, under a representation, becomes the hermitian adjoint of the corresponding operator on Hilbert space. Using that structure, and demanding that we have an identity element $\bf{1}$ in the list of local observables, we define the dimension, $d_i$, of a field $\O_i$ by
\ben
d_i = \frac{1}{2} {\rm sd} \left[ C_{i^\dagger i}^0(x_1,x_2,y) \right]
\label{dim}
\een
where $C_{i^\dagger i}^0$ denotes the coefficient of the identity, $\bf{1}$, in the OPE of $\O^\dagger_i(x_1) \O_i (x_2)$ and ``${\rm sd}$'' denotes the scaling degree \cite{bf} of this distribution, i.e., its rate of divergence as $x_1,x_2 \to y$.

It is useful to introduce the equivalence relation
$u_1 \sim_\delta u_2$
between distributions $u_1$ and $u_2$ on $M^{n+1}$
to mean that the scaling degree of the
distribution $u = u_1 - u_2$ about any
point on the total diagonal $x_1 = \dots = x_n = y$ is $-\delta$.
The properties of the OPE coefficients that we shall give below are ``asymptotic conditions'' in the sense that they hold
at short distances in the sense of $\sim_\delta$ for any choice of $\delta$,
provided the sum(s) in the OPE are carried out to a correspondingly high order. We indicate this by `$\approx$'.

Our axioms for OPEs are as follows:
\begin{itemize}
\item (C1) The OPE coefficients depend locally and covariantly on the background structure $\M=(M,g_{ab},\tau,o)$ \cite{hw1,hw2,bfv}. 
\item (C2) Natural properties hold when one of the fields in the OPE is the identity element $\bf{1}$.
\item (C3) A compatibility condition with $\dagger$ holds. 
\item (C4) $C_{1 \dots n}^k(x_1, \dots, x_n; y)$ satisfies appropriate symmetry/antisymmetry property (depending on the Bose/Fermi nature of the observables) under interchange of $x_i$ and $x_{i+1}$ whenever $x_i$ and $x_{i+1}$ are spacelike related.
\item (C5) The scaling degree of $C_{1\dots n}^j$ on the diagonal satisfies a bound in terms of the dimensions $d_1, \dots, d_n, d_j$ of the fields.
\item (C6) A positivity condition holds, expressing that $\O^\dagger(x_1) \O(x_2)$ appropriately smeared against a positive kernel in $x_1,x_2$ is a positive operator, at least at short distances, giving rise to a positivity property of 
$C_{\O^\dagger \O}^0$.
\item (C7) The OPE coefficients satisfy a ``microlocal spectrum condition'' \cite{bfk}, which will be explained further below. 
\item (C8) The OPE coefficients satisfy an associativity condition, which will be explained further below. 
\item (C9) The OPE coefficients depend on the background structure in a smooth way (see \cite{morettiKhav}).
\end{itemize}
For a more detailed description of these properties, see \cite{HollandsWaldOPE}. Here we only make some comments on (C7) and (C8),
since it is not obvious what a ``spectrum condition'' means in the absence of translation symmetry, nor what an associativity condition means when the product is only understood in an asymptotic sense at short distances. 

A key property of QFT in Minkowski spacetime expressing the stability of the theory is the positivity of energy, i.e., the requirement that the expected total $4$-momentum in any state must be future-directed timelike or null. This is known as the ``spectrum condition.''
In particular, in order for the spectrum condition to hold for the free KG field in Minkowski spacetime, $\phi(x)|0\rangle$ must be a linear combination of one-particle states with timelike or null $4$-momentum. It can be seen that this property holds by looking at the two-point function $\langle 0|\phi(x_1) \phi(x_2)|0\rangle$, which is equal to the coefficient of the identity in (\ref{KGeq}). The Fourier transform of this two-point function is
\begin{equation}
\label{FT2pt}
{\rm FT} \left( \frac{1}{(x_1-x_2)^2 +i\epsilon(x_1^0-x_2^0)} \right) \propto 
\Theta(k^0_1) \delta(k^2_1) \delta^4(k_1+k_2)
\end{equation}
which is non-zero only for past directed null momentum vectors $k^a_2$, associated with a positive energy eigen-ket $|-k_2\rangle$ appearing in $\phi(x_2)|0\rangle$, and future directed null momentum vectors $k^b_1$ associated with a positive energy eigen-bra $\langle k_1|$ appearing in $\langle 0|\phi(x_1)$. This illustrates that the positive energy property is closely related to the Fourier transform properties of the two-point function, which, in turn, is closely related to the coefficient of the identity in the OPE of $\phi(x_1) \phi(x_2)$.

In a general curved spacetime without a stationary symmetry we do not have an unambiguous notion of energy, so we cannot impose a positive total energy condition. Thus, we can at most only impose some sort of local version of the spectrum condition. For the free KG field in curved spacetime, although we do not have a preferred vacuum state, we can consider the identity coefficient, $C^0_{\phi \phi}$, in the OPE of $\phi(x_1) \phi(x_2)$,
Although we do not have a coordinate independent notion of Fourier transform, certain properties of the high frequency behavior of the Fourier-transform of an appropriately localized version of a distribution can be invariantly defined. In particular, there is a well defined notion of  the ``wave front set''
\cite{hormander} of a distribution, which gives a precise characterization of the singular behavior in momentum space $k$ of a distribution localized near point $x$. It is then natural to demand that the wavefront set of $C^0_{\phi \phi} (x_1, x_2)$ take the form
\begin{equation}
        {\rm WF}(C_{\phi \phi}^0) \subset  
        \{ (x_1, k_{1a}, x_2, k_{2b}, y, 0) :
    g_{ab}(x_1) k^a_1 k^b_1  = 0,  k_1^a \nabla_a \tau >0, k_{1}^a \cong -k_2^b \}
    \label{msc}
\end{equation}
 Here $\tau$ is a global time function compatible with the given time orientation, and $k_1^a \nabla_a \tau >0$ corresponds to the $\Theta(k_1^0)$ step function in \eqref{FT2pt}. The notation $\cong$ means that two vectors are tangent to a null geodesic and parallel transported into each other, so the condition $k_{1}^a \cong -k_2^b$ is analogous to the $\delta^4(k_1+k_2)$ momentum conservation delta-function in \eqref{FT2pt}. The condition that 
$g_{ab}(x_1^{}) k^a_1 k^b_1 = 0$ corresponds to the $\delta(k_1^2)$
delta function in \eqref{FT2pt}. The wave front set condition (\ref{msc}) is referred to as a {\it microlocal spectrum condition} on $C^0_{\phi \phi}$. It can be shown that this condition---together with the positivity condition on states---is equivalent to the Hadamard condition \cite{radzikowski}. In other words, the for the free KG field in curved spacetime, the microlocal spectrum condition \eqref{msc} implies that all ``regular'' states $|\Psi\rangle$ must have a two-point function $\langle \Psi | \phi(x_1) \phi(x_2) | \Psi \rangle$ of the Hadamard form. 

For general nonlinear fields, our microlocal spectrum condition (C7) imposes a wave front set condition of the same form as (\ref{msc}) on all OPE coefficients $C_{ij}^k(x_1, x_2,y)$ involving the product of two local fields.
The corresponding condition appropriate for the OPE coefficients for $n>2$ fields is considerably more complicated and is formulated in terms of graphs whose edges are null geodesics labelled by parallel transported null co-vectors. We refer to \cite{HollandsWaldOPE,hollandsQFTCST}
for further explanation.

The associativity condition (C8) imposed on the OPE coefficients is analogous to the usual associativity condition $(a_1 a_2) a_3 = a_1 (a_2 a_3)$ for products in an ordinary finite-dimensional algebra. Its importance in conformally invariant QFTs was first 
emphasized by \cite{Polyakov,Ferraraetal}. In more general QFTs, the OPE is an asymptotic relation, and of course there is a priori no underlying algebra, so the associativity condition must be formulated in terms of the approach of the points
$(x_1, \dots, x_n)$ (where $n>2$) to $y$ at different rates \cite{HollandsWaldOPE}. 

To illustrate this, consider the OPE of three fields $\O_1(x_1) \O_2(x_2)\O_3(x_3)$. We may consider a limit where all 3 points $(x_1, x_2, x_3)$ approach $y$ at the same rate, i.e., we may consider the situation where each of their coordinate displacements from $y$ are of order $\varepsilon$, where $\varepsilon
\to 0$. The asymptotic behavior in this limit is described by the OPE, eq.~(\ref{OPE}), for 3 fields
\ben
\mathcal{O}_{1} (x_1) \O_2(x_2) \O_3(x_3) ) \approx \sum_k C_{123}^k(x_1, x_2, x_3; y) \mathcal{O}_k(y)
\label{assoc1}
\een
Alternatively, we may consider a second situation in which 
$(x_1, x_2, x_3)$ still approach $y$ at rate $\epsilon$ but
$x_1, x_2$ approach each other very rapidly,
say, at rate $\epsilon^2$.
In this second situation, we may view $x_1$ and $x_2$ as first approaching some point $x_4$, and then view $x_3$ and $x_4$ as approaching $y$. In this case, we should be able to first perform for OPE of $\O_1(x_1) \O_2(x_2)$ approaching $x_4$
\begin{equation}
\O_1(x_1) \O_2(x_2) \approx  \sum_l C_{12}^l(x_1, x_2; x_4) \mathcal{O}_l(x_4), 
\end{equation}
We should then be able to plug this expansion into $\O_1(x_1) \O_2(x_2)\O_3(x_3)$ and then perform an OPE of $\O_l(x_4) \O_3(x_3)$
\ben
\O_l(x_4) \O_3(x_3) \approx \sum_m C_{l3}^m(x_4, x_3; y) \mathcal{O}_m(y)
\label{assoc3}
\een
The associativity condition asserts that the final results should be the same. This gives a relationship between the OPE coefficients $C_{123}^k(x_1, x_2, x_3; y)$ for three fields and sums of products of OPE coefficients for two fields. In order to formulate this condition precisely and generalize it to the case of OPE coefficients for $n$ fields, one must introduce the notion of ``merger trees'' that describe the hierarchical approach of $n$ points to coincidence and apply the above defined notion of $\sim_\delta$ for each merger tree.
We refer the reader to \cite{HollandsWaldOPE} for the details of how this can be done. We note that the associativity condition provides a separate relation for each merger tree.

Finally, we note that OPEs can be used to obtain local relations on field observables as follows. Suppose a local field observable $\O$ has the property that all terms in the OPE expansion of $\O^\dagger(x_1) \O(x_2)$ have strictly negative scaling degree. Then we demand that $\O = 0$. In this way, given the OPE coefficients, it can be checked as to whether various local PDE relations hold among the field observables.

In summary, we believe that the formulation of QFT should be based on its OPEs and that the specification of a full list of OPE coefficients $C_{1 \dots n}^k(x_1, \dots, x_n; y)$ that satisfy the above axioms should be viewed as a specification of the QFT.

\section{Flow relations for OPEs in the coupling parameters}
\label{sec:action}

Whether one takes the view that we have just advocated that OPEs are fundamental to the formulation of QFT, or one merely takes the view that OPEs are a convenient and useful tool in QFT, the task remains to find the OPEs of a given QFT of interest. Except for ``superrenormalizable''
QFTs whose OPE coefficients only depend on finitely many perturbation orders, it is not realistic to expect
that there should be a simple, explicit, algorithm for constructing the OPE coefficients of a general nonlinear QFT. However, if the nonlinear theory arises from a Lagrangian of the form
\ben
L = L_{\rm F} + \lambda \O
\een
where $L_{\rm F}$ is the Lagrangian of a linear field theory (e.g., KG theory) and $\O$ is some nonlinear observable (e.g., $\O = \phi^4$), then one could apply perturbative methods in the coupling parameter $\lambda$ to obtain perturbative corrections to the OPE coefficients of the free theory. Standard perturbative methods would enable one to obtain, in principle, corrections to the OPE coefficients to any finite order in $\lambda$. However, such methods would be extremely cumbersome and, of course, issues of convergence of the perturbative series would remain in general.

A much more promising approach arises from considering how the OPE coefficients change under a change of the coupling parameter. As a simple, toy example of this, consider the Green's function $G_{\rm E}(x_1, x_2; m^2)$ for the operator 
\begin{equation}
K=-\delta^{ab}\partial_{a}\partial_{b}+m^{2},\label{eq:Euclidean K}
\end{equation}
where $\delta_{ab}$ denotes the flat Euclidean metric on ${\mathbb{R}^{4}}$ and $m^2 > 0$. This Green's function will play the role of the coefficient of $\bf{1}$ in the OPE of $\phi(x_1) \phi(x_2)$ in a Euclidean version of massive KG QFT. One may ask how $G_{\rm E}(x_1, x_2; m^2)$ changes under a change of mass parameter $m^2$. It is not difficult to show that the following relation holds \cite{KW}:
\begin{equation}
\frac{\partial}{\partial m^{2}}G_{E}(x_{1},x_{2};m^{2})=-\int d^{4}yG_{E}(y,x_{1};m^{2})G_{E}(y,x_{2};m^{2}).\label{Gflow}
\end{equation}
Thus, if we knew $G_{\rm E}(x_1, x_2; m^2)$ at some value $m^2 = m^2_0$, we could, in principle, 
integrate the differential equation (\ref{Gflow}) to obtain $G_{\rm E}(x_1, x_2; m^2)$ at other values of $m^2$. Of course, in this case, solving (\ref{Gflow}) would be a much more difficult task than obtaining $G_{\rm E}(x_1, x_2; m^2)$ by direct techniques. However, if we did not have such direct techniques but happened to know $G_{\rm E}(x_1, x_2; m_0^2)$, then (\ref{Gflow}) would provide a potentially valuable tool for obtaining $G_{\rm E}(x_1, x_2; m^2)$ for $m^2 \neq m_0^2$. 

It has been shown in \cite{Holland:2014ifa,Holland:2015tia,HollandsAction} that ``flow relations'' in the coupling parameter analogous to (\ref{Gflow}) exist for the OPE coefficients in Euclidean versions of renormalizable theories.
In the case of $\lambda \phi^4$-theory, the local fields appearing the OPEs are of the form $\mathcal{O}_i = \partial^{\alpha_1} \phi \cdots \partial^{\alpha_N} \phi$, where each $\alpha_i$ is a multi-index. 
Since $\phi$ is a bosonic field, the labels $i$ in this theory are in correspondence with the {\em un}-ordered tuples $\{ \alpha_1, \dots, \alpha_N\}$ of multi-indices. The flow equation for the OPE coefficients takes the form \cite{Holland:2014ifa,Holland:2015tia,HollandsAction}
\begin{multline}
\label{eq:FE}
    \partial_\lambda C^j_{1 \dots n} (x_1, \dots, x_n; z) = 
    \int d^4 y \, \chi(y,z) \Bigg[
C_{\phi^4 1 \dots n}^j(y,x_1,\dots,x_n;z) \, \\
-\sum_{k=1}^n \sum_{d_\O \le d_k} C_{\phi^4 k}^\O(y, x_k;x_k) C^j_{1
\dots k \!\!\! / \O \dots n}(x_1, \dots, x_n;z) \, \\
-\sum_{d_\O<d_j} C^\O_{1 \dots n}(x_1, \dots x_n; z) C^j_{\phi^4 \O}(y,z;z)
    \Bigg]
\end{multline}
Here $\chi(y,z)$ is a suitable cutoff function, which can be taken to be
the characteristic function of the set $\{|y-z| \le L\}$ for some infra-red scale $L$, which reduces the integration domain to a compact set. The sums are taken over all observables $\O$ satisfying the indicated restrictions on their dimension $d_{\O}$.
In perturbation theory, the dimension $d_i$ is to leading order the ``engineering dimension'' (i.e. that assigned according to dimensional analysis by its appearance in the Lagrangian) of a given composite field, plus an ``anomalous'' correction of order $\lambda$ that would be extractable 
from the 2-point OPE coefficient $C_{i^\dagger i}^0$ (see \eqref{dim}). 

It can be shown \cite{Holland:2014ifa} that the above flow equations can be used do define the OPE coefficients as formal power series in $\lambda$ taking as the initial condition for the subsequent $\lambda$-integrations the known values for the OPE coefficients in the underlying free QFT defined by $\lambda = 0$. Thus, the flow equation (\ref{eq:FE}) together with these initial conditions at $\lambda = 0$ enables one to formally obtain the OPE coefficients of the interacting theory in a direct manner, thereby bypassing the standard tools of perturbative QFT such as Feynman diagrams, path integrals, or renormalization group flow equations. 
Furthermore, as a consequence of the flow equation, the interacting field OPE coefficients automatically satisfy the associativity axiom (C8) as well as Euclidean counterparts of other axioms of the previous section.

Unlike most traditional perturbative approaches, equation \eqref{eq:FE} does not require a Lagrangian formulation. All that is needed is an initial point of the flow at which we know all OPE coefficients of the theory. Thus, the flow equation \eqref{eq:FE} is in principle also useful to study deformations of explicitly soluble models other than free fields such as the conformally invariant so called ``minimal models'' \cite{bpz} in 2-dimensional Euclidean space. For a discussion of  flows that remain conformal, see \cite{HollandsAction}.

Although (\ref{eq:FE}) provides satisfactory flow relations in flat Euclidean space, one would like to have such flow relations in curved Lorentzian spacetimes. A number of difficulties arise in generalizing (\ref{eq:FE}) to curved Lorentzian spacetimes, and in order to overcome these, the following modifications must be made \cite{KW}. First, in the Lorentzian case, flow relations do not exist for the coefficients arising in the OPE expansion of the ordinary products $\O_1 (x_1) \cdots \O_n (x_n)$. Instead, one must work with the coefficients arising in the OPE expansion of time-ordered products ${\rm T}[\O_1 (x_1) \cdots \O_n (x_n)]$. Time-ordered products have renormalization ambiguities on the ``diagonals'', i.e., at coincidence of the points, so the use of time-ordered products has the potential to lead to significant further complications. Fortunately, however, to define the flow relations, it suffices to consider only the case where 
$\{x_1, \dots, x_n\}$ are pairwise distinct, in which case time-ordered products are well defined. Although the integration over $y$ in the flow equation \eqref{eq:FE} includes coincident points, the associativity condition (C8) guarantees that 
the particular combination of terms in square brackets in equation \eqref{eq:FE} vanishes to a sufficiently high order at 
coincident points that the integrand can be defined by continuity.

Second, in Euclidean space the cutoff function $\chi$ appearing in \eqref{eq:FE}---which is needed to guarantee convergence of the integral---can be chosen in a manner compatible with rotational invariance. However, even in Minkowski spacetime, no choice of cutoff function can be Lorentz invariant. Thus, if flow relations were defined by a direct analog of \eqref{eq:FE}, they would yield non-Lorentz invariant OPE coefficients in Minkowski spacetime and would yield non-covariant OPE coefficients in a curved Lorentzian spacetime. The solution to this difficulty \cite{KW} is to add ``counterterms''---which are canonically determined by $\chi$ and the OPE coefficients themselves---to the integrand on the right side of \eqref{eq:FE} so as to restore Lorentz covariance.

A final difficulty is that an integration over any finite region in $y$ of quantities local in $(y,x_1, \dots, x_n,z)$ as appears in \eqref{eq:FE} cannot produce a result that is local in $(x_1, \dots, x_n,z)$ in the sense required by axiom (C1). The solution to this difficulty \cite{KW} is to replace the quantities appearing in the integrand of \eqref{eq:FE} by their Taylor expansion about $z$, carried out to sufficiently high order to accommodate the desired scaling degree in the flow relation. 

Thus, an analog of the flow relations \eqref{eq:FE} can be given in curved Lorentzian spacetimes, although it is considerably more complicated and cumbersome than in the Euclidean case. Nevertheless, these flow relations can be viewed as giving 
a closed system of equations for the OPE coefficients of the interacting theory that, formally, should determine them given the initial values defined by the free theory.

\section{OPEs in conformal QFTs and conformal bootstrap}
\label{sec:cft}

From a practical viewpoint, the OPE as a mathematical framework for QFT has had its most striking success by far for conformal QFTs with a conserved stress energy tensor on $2$-dimensional Minkowski spacetime. This line of investigation started with the paper \cite{bpz} which, in essence, 
applied the basic ideas of \cite{Polyakov,Ferraraetal} to $D=2$ dimensions, leading to large classes of exactly solvable QFTs in the continuum. More recently, significant progress has also been made in $D > 2$ spacetime dimensions, and we now review some of the basic ideas.

The success in $D=2$ dimensions can be traced back to the existence of an infinite-dimensional group of symmetries of such QFTs---namely, Virasoro symmetry---that commutes with the OPE. This considerably reduces the amount of data that one needs to determine in order to have access to the full OPE. A widely used modern mathematical formulation of the OPE in conformal QFTs in $D=2$---taking full advantage of this extra structure, and essentially boiling down to an algebraic structure in the traditional sense---is the setting of ``vertex operator algebras'' (see e.g. \cite{vertex2}).
The basic idea here---and in any other situation wherein a group (or algebra) is acting on the space of local operators as a symmetry of the OPE---is that one may arrange the local fields into multiplets transforming under the irreducible representations of this group. The OPE coefficients must respect this multiplet structure. The multiplets contain ``more'' local operators if the symmetry group is large, thereby placing more restrictions on the OPE. In the extreme case of rational conformal QFTs in $D=2$, there are only {\em finitely many} multiplets under the Virasoro algebra, so the algebraic data determining the OPE coefficients can be encoded in finitely many ``fusion rules.'' Furthermore, these fusion rules to a certain extent can be classified algebraically within the theory of ``modular tensor categories'', see e.g. \cite{mtc,bakalov,Schweigert}. 

In the case of conformal QFTs in $D \ge 3$ spacetime dimensions, there are  infinitely many multiplets (now under the ordinary finite dimensional conformal algebra $\mathfrak{o}(D,2)$), but the structure is still sufficiently restrictive that it provides a significant simplification of the OPE. Roughly speaking, each multiplet consists of a ``primary'' local operator, $\mathcal{O}$, together with its ``descendants'', $\{ \partial_{a_1} \dots \partial_{a_k} \mathcal{O}\}$, where the partial derivatives $\partial_a$ correspond from a representation theoretic viewpoint to generators $iP_a$. 

The existence of multiplets implies that we can restrict consideration to the OPE of primary fields only. Furthermore, the OPE of two primary operators $\mathcal{O}_i, \mathcal{O}_j$ has the special structure
\cite{Ferraraetal,schroer,Dolan1,Dolan2}
\begin{equation}
\label{CFTOPE}
    \mathcal{O}_i(x_1) \mathcal{O}_j(x_2) = 
    \sum_k \frac{f_{ijk}}{|x_1-x_2|^{d_i+d_j-d_k}} P_{ij}^k(x_1-x_2; y, \partial_y) \mathcal{O}_k(y) ,  
\end{equation}
where $\mathcal{O}_k$ also is primary. The complex numbers $f_{ijk}$ are called ``structure constants'' and the non-negative numbers $d_i$ are the dimensions of the fields \eqref{dim}. The differential operators $P_{ij}^k$ are entirely kinematical, i.e. determined by the representation theory of the conformal algebra $\mathfrak{o}(D,2)$, which also is used to assign a representation theoretic label called ``spin'', $\ell_i$, to each field. 
Thus, the OPE of the given conformal QFT is determined entirely in terms of its so called ``conformal data'', $\{d_i, \ell_i, f_{ijk}\}$, which by itself is already a huge simplification over the infinite number of OPE distributions without a priori fixed functional form that we would normally encounter in non-conformal QFTs. 

Another key simplification in conformal QFTs on Minkowski spacetime is that the OPE  is expected to {\em converge}\footnote{The OPE also converges to any finite order in perturbation theory in Euclidean space \cite{kopper} though this 
is no longer expected to be the case in general curved spacetimes or at the non-perturbative level.} \cite{mack1976,pappadopuloetal2012} -- rather than just being an {\em asymptotic} scaling relation associated with different merger trees in the sense indicated in section \ref{sec:axioms} and discussed in more detail in \cite{HollandsWaldOPE}. As a consequence, the associativity condition (C8) can be expressed in terms of convergent series and equality signs. In other words, the kinds of asymptotic conditions implied by eqs.~(\ref{assoc1})-(\ref{assoc3}) above would now be replaced by {\bf exact equalities} when the points are separated by a sufficiently small but {\em finite} $\varepsilon>0$. 

Together the above features of conformally invariant QFTs in $D \ge 3$ dimensional Minkowski spacetime can be combined to give non-trivial a priori information about the conformal data. This idea is the basis of the so-called ``conformal bootstrap'' approach, the basic idea of which goes back to \cite{Polyakov,Ferraraetal}. Beginning with \cite{Ratazzi}, the practical efficiency of this approach, suitably formulated, has been appreciated and applied very widely. We now very briefly sketch this approach; see e.g. the reviews by \cite{rychkov2,polandetalreview,Simmons-Duffin:2016gjk} for more detail.

First, under the assumption of conformal invariance, the OPE coefficients 
for the identity operator (corresponding to the upper label `$0$') in a general conformally invariant QFT in 
$D\ge 3$ dimensions are:
\begin{eqnarray}
C_{12}^{0}(x_1,x_2;y) &=& \frac{\delta_{12}}{|x_{12}|^{2d_1}} \\
C_{123}^{0}(x_1,x_2,x_3;y) 
&=& 
\frac{f_{123}}{
|x_{12}|^{d_1+d_2-d_3}
|x_{23}|^{d_2+d_3-d_1}
|x_{13}|^{d_1+d_3-d_2}
}
\end{eqnarray}
where we have written $x_{ij}=x_i-x_j$, and where 
the expression is independent of $y$ since we are considering the 
identity operator $\bf 1$. These expressions are equivalent to formulas for the
2- and 3-point functions in the vacuum state, provided that, if necessary, we subtract a suitable multiple of $\bf 1$ from each operator to ensure that  
it has vanishing 1-point function.
Here, for simplicity, we have taken all three primaries to be scalar and real---in general the conformal data must be enhanced by the spin $\ell_\O$ associated with each operator---and we have analytically continued the OPE coefficients from the Lorentzian to Euclidean domain, as is possible since they are boundary values of analytic functions by the analytic version of the microlocal spectrum condition (C7) and well-known connections between the analytic version of the wave front set and boundary values of analytic functions \cite{hormander}. 

Useful constraints on the OPE coefficients arise from applying the associativity condition to the OPE coefficients $C_{1234}^{0}$ associated with four fields $\O_1, \dots, \O_4$ and the identity field $\bf 1$
corresponding to the upper index `$0$': We can consider the merger tree where $x_{12}$ and $x_{34}$ are of order $\varepsilon^2$ but $x_{13}$ is of order $\varepsilon$ (the ``$s$-channel''). In this case, we can first perform OPEs in $x_1$ and $x_2$ and in $x_3$ and $x_4$ before combining these expansions to get an OPE for $ \O_1 (x_1) \O_2(x_2) \O_3(x_3) \O_4(x_4)$ in the manner described near the end of section \ref{sec:axioms}. Alternatively, we can obtain another expression by carrying out the same steps for the merger tree where $x_{13}$ and $x_{24}$ are of order $\varepsilon^2$ but $x_{12}$ is of order $\varepsilon$ (the ``$t$-channel''). For conformal QFTs, each of these OPE expressions has a finite domain of convergence. Furthermore, these domains of convergence have a finite overlap as follows: Define the conformally invariant cross-ratios
\begin{equation}
    u = \frac{|x_{12}|^2 |x_{34}|^2}{|x_{13}|^2 |x_{24}|^2}, \quad
    v = \frac{|x_{14}|^2 |x_{23}|^2}{|x_{13}|^2 |x_{24}|^2}
\end{equation}
Then we have convergence of both the ``s'' and ``t'' channel expressions when $u < 1$ and $v < 1$. Associativity then yields a relation of the form
\begin{equation}
\label{crossingsym}
    \sum_\O f_{12\O} f_{34\O} \frac{G^{12,34}_\O(z,\bar z)}{|z|^{d_1+d_2}} = 
    \sum_\O f_{13\O} f_{24\O} \frac{G^{13,24}_\O(1-z,1-\bar z)}{|1-z|^{d_1+d_3}}, 
\end{equation}
where $z$ is the complex parameter defined by $u=|z|^2, v=|1-z|^2$. Here the sum is over all primary fields $\O$ of the theory, and each $G^{ij,kl}_\O(z,\bar z)$ denotes an appropriately normalized so-called ``conformal block'' \cite{Dolan1,Dolan2,Dolan4}. The conformal blocks\footnote{Here, we restricted ourselves to scalar operators; in the general case certain bases of tensor structures must be considered, leading to the 
consideration of ``spinning conformal blocks'', see e.g. \cite{Costa, Costa:2016xah}.} depend only on the conformal dimensions  $\{d_i\}$ and spins $\{\ell_i\}$ of the operators in the theory and are computable by group-theoretical or analytical methods, see e.g. \cite{Dolan1,Dolan2,Dolan4,Kravchuk:2017dzd,penedones,Karateev:2017jgd}.
Thus, \eqref{crossingsym} reduces to a condition on the conformal data $\{d_i, \ell_i, f_{ijk}\}$. 

In practice, these conditions are used to obtain bounds on the conformal data under suitably chosen extra structural hypotheses, such as field content, number of fields with ``low'' conformal dimensions, conservation laws, unitarity, (super-)symmetry, or other a priori information. 
Starting with \cite{Ratazzi}, a common scheme implementing this idea has been to take arbitrary derivatives with respect to $z,\bar z$ of the ``crossing relations'' \eqref{crossingsym}, and to evaluate them at $z=1/2$. 
In practice, some variations of this method can be more practical depending on the situation; see e.g. \cite{Caracciolo:2009bx,Poland:2011ey}, or the general references \cite{rychkov2,polandetalreview,Simmons-Duffin:2016gjk} for more detail. Combined, these methods turn out to be extremely efficient and versatile. The 3D Ising model at the critical point is a showcase example of the method, which has given spectacularly precise results on the conformal dimensions of ``low lying'' operators (see e.g. \cite{ElShowk:2012ht,Kos}).

\section{The PCT theorem in curved spacetime}
\label{sec:theorems}

In this section, we use the example of the PCT theorem to illustrate how general results on QFT can be formulated and proven within the OPE framework presented in section \ref{sec:axioms}. The PCT theorem in Minkowski spacetime states that in any QFT satisfying the Wightman axioms, there exists an anti-unitary map on the Hilbert space (or, in the algebraic setting, an anti-linear $*$-isomorphism on the field algebra) that relates each field at point $x$ to its conjugate field at $-x$, where ``$-x$'' denotes the action of the $PT$ isometry on Minkowski spacetime (where $P$ denotes the parity isometry and $T$ denotes the time reversal isometry). The proof of the PCT theorem in Minkowski spacetime mainly relies on the facts that (i) the spectrum condition implies that $n$-point functions in the vacuum state have an analyticity property in complex $t$ and (ii) the isometry $PT$ lies in the connected component of the identity in the complexified Lorentz group. It might seem that there could not be any meaningful generalization of either the formulation or proof of the PCT theorem to general curved spacetimes, since, in particular, a general curved spacetime admits neither a $PT$ symmetry nor a Lorentz group symmetry. Nevertheless, we will now outline how this can be done.

Instead of attempting to relate fields to their conjugate fields under the action of some parity and time reversing diffeomorphism on a given spacetime, we instead seek a relationship between the QFT on the background structure $\M=(M,g_{ab},\tau,o)$ and the QFT on the background structure $\bar \M=(M,g_{ab},-\tau,o)$, i.e., $\bar \M$ is the same spacetime $(M, g_{ab})$ but with the opposite choice of time and space orientations (and, thus, the same choice of spacetime orientation $o$). The general curved spacetime version of the PCT theorem states that if the OPE coefficients $C_{1\dots n}^j$ of the QFT satisfy the axioms of section \ref{sec:axioms}, then (restricting, for simplicity to $D=4$ spacetime dimensions) we have \cite{HollandsWaldOPE}
\begin{equation}
C_{1\dots n}^j[\bar \M] \approx 
\overline{C_{1^\dagger\dots n^\dagger}^{j^\dagger}[\M]} \,
 i^{-F_j+F_1+\dots+F_n} (-1)^{-U_j+U_1+\dots+U_n}
\end{equation}
Here $C_{1^\dagger\dots n^\dagger}^{j^\dagger}$ denotes the OPE coefficient of the corresponding conjugate fields (with the overline on this quantity denoting complex conjugation), $F=0, 1$ is the Bose/Fermi parity of the local field $\O$, and $U$ is the net number of unprimed spinor indices of $\O$. Thus, our formulation of the PCT theorem in curved spacetime relates the OPE coefficients of fields on $\M$ to the OPE coefficients of the conjugate fields on $\bar \M$. In Minkowski spacetime, we can combine this result with the $*$-isomorphism between the field algebras on $\M$ and $\bar \M$ given by the $PT$ isometry to obtain a relation purely between the OPE coefficients on $\M$, which corresponds to the usual formulation of the PCT theorem.

It is instructive to give a sense of how this result can be proven, since a general curved spacetime need not be analytic in $t$, so the kind of analytic continuation of $n$-point functions to complex $t$ done in Minkowski spacetime cannot be straightforwardly carried out in a general curved spacetime. Nevertheless, one can proceed by Taylor expanding the metric up to any arbitrary given order $N$ about an arbitrary point $y$ in Riemannian normal coordinates $x^\mu$ centered at $y$,
\begin{equation}
    g_{\mu\nu}(x^\sigma) = \sum_{k=0}^N p_{\mu\nu\sigma_1 \dots \sigma_k}(y) x^{\sigma_1}
    \dots x^{\sigma_k} + O(|x|^{N+1}), 
\end{equation}
where each $p_{\mu\nu\sigma_1 \dots \sigma_k}$ is constructed from the metric, the Riemann tensor, $R_{\alpha\beta\gamma\delta}$, and its covariant derivatives at point $y$. Using this Taylor expansion and using the covariance and scaling properties of the OPE coefficients, one can then show that
\begin{equation}
    C_{1\dots n}^j(x_1,\dots,x_n,y) \sim_\delta
    \sum_{I} q^I(y) W_{I1\dots n}^j(x_1,\dots,x_n),
\label{PCT}
\end{equation}
where $q^I$ are curvature polynomials of the sort just described, 
and each $W_{I1\dots n}^j(x_1,\dots,x_n)$ is a Lorentz invariant 
$n$-point distribution in the tangent space at $y$, which may be identified with a Lorentz invariant distribution on Minkowski spacetime. Here $\delta$ is 
a chosen accuracy to which the two sides are equal in the coincidence limit and the finite sum over $I$ includes sufficiently many terms depending on this choice of $\delta$. The microlocal spectrum condition (C7) then implies an analyticity property of $W_{I1\dots n}^j(x_1,\dots,x_n)$ analogous to the analyticity property of $n$-point functions in the vacuum state in Minkowski quantum field theory. One may then apply the arguments used in the proof of the PCT theorem in Minkowski spacetime to derive relations on $W_{I1\dots n}^j(x_1,\dots,x_n)$ that yield \eqref{PCT}. We refer the reader to \cite{HollandsWaldOPE} for further details as well as a formulation and proof of the spin-statistics theorem in a general curved spacetime.

\section{Conclusions}

As we have illustrated in this review, operator product expansions play a fundamental role in the structure of quantum field theory and provide important tools for the elucidation of the properties of quantum fields.

\end{document}